\documentclass[12pt]{article}
\usepackage[dvips]{epsfig}
\usepackage[T1]{fontenc}
\usepackage[latin1]{inputenc}
\usepackage{graphicx}
\usepackage[english]{babel}
\usepackage{amsmath}
\usepackage{amssymb}
\usepackage{amsfonts}
\usepackage[T1]{fontenc}
\usepackage{color}
\usepackage{babel}
\usepackage{verbatim}
\usepackage[unicode=true,pdfusetitle,bookmarks=true,bookmarksnumbered=false,bookmarksopen=false,
 breaklinks=false,pdfborder={0 0 1},backref=false,colorlinks=true]{hyperref}
\hypersetup{linkcolor=blue,citecolor=blue}
\makeatletter
\usepackage[dvips]{epsfig}
\usepackage[T1]{fontenc}
\usepackage{color}
\usepackage{pdflscape}

\begin{document}

\title{\bf Nonlocal Ernst Equations}
\author{Metin G{\" u}rses$$\thanks{%
email: gurses@fen.bilkent.edu.tr},
\\{\small Department of Mathematics, Faculty of Sciences, Bilkent University, 06800 Ankara, Turkey}}

%\makedate
\maketitle

\begin{abstract}
We show that the Ernst equations for stationary axially symmetric Einstein-Maxwell and Einstein - N-abelian Yang-Mills field equations have local and  nonlocal reductions. Among these reduced equations the nonlocal Ernst equations are new. We show that a class of these field equations admit reflection symmetry without requiring the asymptotical flatness.

\end{abstract}

%%%%%%%%%%%%%%%%%%%%%%%%%%%%%%%%%%%%%%%%%%%%%%%%%%%%%%%%%%%%%%%%%
\section{Introduction}
In the last decade there  have been some new reductions invented in the integrable nonlinear partial differentiable equations which was initiated by the works of Ablowitz and Musslimani \cite{ab-muss1}-\cite{ab-muss3}. This is a kind of reduction where in the resulting equations not only the point $(t,x)$ is involved but also the space reversal or time reversal or time and space reversal points are involved. Such reductions are called nonlocal reductions. More studied examples are in nonlocal NLS \cite{ab-muss1}-\cite{gur5}, nonlocal KdV \cite{gur4} and nonlocal mKdV equations \cite{gur5},\cite{ab-muss2},\cite{ab-muss3}. Soliton solutions of these new equations have been found by the metohd of inverse scattering method, Darboux transformation and Hirota bi;linear method. It ha been later understood that if the system of equations admit scale symmetry then this system admits nonlocal reductions \cite{gur3}.

Einstein-Maxwell field equations for spacerimes admitting two commuting Killing vector field reduce two coupled complex equations \cite{solbook} known as the Ernst equations \cite{ernst1},\cite{ernst2}. Local reduction of these equations were considered before
\cite{ernst2}. In this work we shall consider both local and nonlocal reductions of these equations for stationary axially symmetric spacetimes. We obtain one new equation from a local reductions and two new equations from two different nonlocal reductions.

In the study of stationary axially symmetric vacuum  Einstein field equations, a class of solutions admit reflection symmetry \cite{kordas}- \cite{meinel}. In this subclass the Ernst complex function satisfies  $ \bar{\xi}(\rho,z)= \xi(\rho,-z)$.  Reflection symmetry exists also in stationary axially symmetric electrovacuum field equations \cite{ernst3}, \cite{ernst4}.  We verify this fact by the use of the nonlocal Ernst equations. We show that, in stationary axially symmetric Einstein-Maxwell and Einstein - N-abelian Yang-Mills field equations there exist a class of solutions possessing the reflection symmetry.

\section{Einstein-Maxwell Ernst Equations}

Let the spacettime  be stationary and axially symmetric, then the line element takes the form
\begin{equation}
ds^2=f^{-1}\,[e^{2 \gamma}(d\rho^2+dz^2)+\rho^2 \,d \phi^2]-f\, (dt-w d\phi)^2
\end{equation}
where $f$, $\gamma$ and $w$ are differentiable functions of $\rho$ and $z$. Let $A_{t}$ and $A_{\phi}$ be the components of the elecetromagnetic 4-potential. Then after playing with the Einstein Maxwell field equations and defining two complex potentials \cite{ernst2} one obtains

%Einstein-Maxwell field equations in a spacetime admitting two noncommuting Killing vector fields can be arranged so that essential part of the field %equations take the following form

\begin{eqnarray}
&&({\cal{E}+\bar{\cal{E}}}+2 \Phi \, \bar{\Phi})\, \nabla^2\, {\cal{E}}=2(\vec{\nabla} {\cal{E}}+ 2 \bar{\Phi} \vec{\nabla}\, \Phi)\, \cdot \vec{\nabla} \cal{E}, \\
&&({\cal{E}+\bar{\cal{E}}}+2 \Phi \, \bar{\Phi})\, \nabla^2\, \Phi=2(\vec{\nabla} {\cal{E}}+ 2 \bar{\Phi} \vec{\nabla}\, \Phi)\, \cdot \vec{\nabla} \Phi,
\end{eqnarray}
where $\cal{E}$ and $\Phi$ are the gravitational and electromagnetic complex potentials. The differential operators $\nabla^2$ and $\nabla$ are defined on the 2-surface orthogonal to the Killing directions. There are two distinct cases: (i) Both of the Killing vectors are spacelike (Colliding gravitational plane wave geometry), (ii) One of the Killing vector is timelike and the other one is spacelike (Stationary axially symmetric spacetime geometry). The above equations are called the Ernst equations \cite{ernst1},\cite{ernst2} for the Einstein-Maxwell field equations. To make the above equations more symmetrical we let

\vspace{0.5cm}
\noindent

\begin{equation}
{\cal{E}}=\frac{\xi-1}{\xi+1},~~\Phi=\frac{\eta}{\xi+1},
\end{equation}
then we obtain

\begin{eqnarray}
&&(\xi \bar{\xi}+\eta\, \bar{\eta}-1)\, \nabla^2 \, \xi=2 (\bar{\xi}\, \vec{\nabla} \xi+ \bar{\eta}\, \vec{\nabla} \eta)\, \cdot \vec{\nabla} \xi, \label{denk1}\\
&&(\xi \bar{\xi}+\eta\, \bar{\eta}-1)\, \nabla^2 \, \eta=2 (\bar{\xi}\, \vec{\nabla} \xi+ \bar{\eta}\, \vec{\nabla} \eta)\, \cdot \vec{\nabla} \eta, \label{denk2}
\end{eqnarray}
In this work we focus on the stationary axially symmetric spacetimes. Hence
both $\xi$ and $\eta$ are functions of the prolate spheroidal coordinates $x$ and $y$ and the differential operators, in this coordinate system take the form

\begin{eqnarray}
&&\nabla^2 \Psi=\frac{1}{x^2-y^2}\, \left[ \frac{\partial}{\partial x} (x^2-1)\frac{\partial \Psi}{\partial x} +\frac{\partial}{\partial y} (1-y^2)\frac{\partial \Psi}{\partial y} \right], \\
&& \vec{\nabla} \Psi \cdot \vec{\nabla} \Phi=\frac{1}{x^2-y^2}\, \left[(x^2-1)\, \frac{\partial \Psi}{\partial x}\, \frac{\partial \Phi}{\partial x}\,+(1-y^2)\, \frac{\partial \Psi}{\partial y}\, \frac{\partial \Phi}{\partial y} \right]
\end{eqnarray}
for any functions $\Psi$ and $\Phi$.
Then the set of equations (\ref{denk1} ) and (\ref{denk2}) admits the following local and nonlocal reductions:

\vspace{0.5cm}
\noindent
{\bf Local Reductions}:  There are two local reductions

\vspace{0.5cm}
\noindent
{\bf (1).} $\eta=\alpha+\beta \xi$. (a) $\beta \ne 0$. The reduced equation can further be related to the vacuum Ernst equation by letting

\begin{eqnarray}
&&\xi=\gamma \sqrt{w}\, E_{0}-\frac{\alpha \gamma \bar{\beta}}{\gamma \bar{\gamma}+\beta \bar{\beta}}, \\
&&\eta=\beta \sqrt{w}\, E_{0}+\frac{ \alpha \gamma \bar{\gamma}}{\gamma \bar{\gamma}+\beta \bar{\beta}},
\end{eqnarray}
where
\begin{equation}
w=\frac{1-\alpha \bar{\alpha}}{\gamma \bar{\gamma}+\beta \bar{\beta}}+\frac{\alpha \bar{\alpha} \gamma \bar{\gamma}}{(\gamma \bar{\gamma}+\beta \bar{\beta})^2}
\end{equation}
and $E_{0}$ satisfies the vacuum Ernst equation
\begin{equation}
(E_{0} \bar{E_{0}}-1)\, \nabla^2 E_{0}=2 \bar{E_{0}}\, ( \vec{\nabla} E_{0}\, \cdot \vec{\nabla} E_{0}) \label{ernst}
\end{equation}
Kerr solution corresponds to $E_{0}=p \,x+i\, q \,y$ with $p^2+q^2=1$.\\
(b) $\beta=0$. Letting $\xi=\rho \,E_{0}$. Then $E_{0}$ satisfies the above Ernst equation (\ref{ernst}) with $\rho \bar{\rho}+\alpha \bar{\alpha}=1$.

\vspace{0.5cm}
\noindent
{\bf (2).} $\eta=\alpha+\beta \bar{\xi}$. (a) $\beta \ne 0$. Compatibility conditions require $\alpha=0$ and $\beta \bar{\beta}=1$. Then the reduced equation takes the form

\begin{equation}
\vec{\nabla} \cdot \left( \frac{{\vec{\nabla}} \xi} {2 \xi \,\bar{\xi}-1} \right)=0 \label{eq1}
\end{equation}
(b) $\beta = 0$.
Letting $\xi=\rho \,E_{0}$. Then $E_{0}$ satisfies the above Ernst equation (\ref{ernst}) with $\rho \bar{\rho}+\alpha \bar{\alpha}=1$.

\vspace{0.5cm}
\noindent
{\bf Nonlocal Reductions}:  Let
\begin{equation}
\xi^{\varepsilon}=\xi(\varepsilon_{1} x, \varepsilon_{2} y), ~~\eta^{\varepsilon}=\eta(\varepsilon_{1} x, \varepsilon_{2} y), ~~~(\varepsilon_{1})^2=(\varepsilon_{2})^2=1
\end{equation}

\vspace{0.5cm}
\noindent
{\bf (1)}.  Let $\eta(x,y)=k \, \xi(\varepsilon_{1} x, \varepsilon_{2} y)$. Then (\ref{denk1}) and (\ref{denk2} reduce to a single equation

\begin{equation} \label{eq4}
(\xi \bar{\xi}+\xi^{\varepsilon}\, \bar{\xi}^{\varepsilon}-1)\, \nabla^2 \, \xi=2 (\bar{\xi}\, \vec{\nabla} \xi+ \bar{\xi^{\varepsilon}}\, \vec{\nabla} \xi^{\varepsilon})\, \cdot \vec{\nabla} \xi,
\end{equation}
 we call this equation as the first nonlocal Ernst equation.

\vspace{0.5cm}
\noindent
{\bf (2)}.  Let ~~$\eta(x,y)=k \, \bar{\xi}(\varepsilon_{1} x, \varepsilon_{2} y)$. Then  (\ref{denk1}) and (\ref{denk2} reduce to the following equation

\begin{equation}\label{eqn5}
(\xi \bar{\xi}+\xi^{\varepsilon}\, \bar{\xi}^{\varepsilon}-1)\, \nabla^2 \, \xi=2 (\bar{\xi}\, \vec{\nabla} \xi+ \xi^{\varepsilon}\, \vec{\nabla} \bar{\xi}^{\varepsilon})\, \cdot \vec{\nabla} \xi,
\end{equation}
where $k \bar{k}=1$ for both cases. This equation is  the second nonlocal Ernst equation. Both of the first and second nonlocal Ernst equations are new.

\section{$N$-Abelian Yang-Mills Coupled to Gravity}

\vspace{0.5cm}
\noindent
The above coupled equations (\ref{denk1}) and (\ref{denk2}) can further be generalized to Einstein-N Abelian gauge field equations \cite{gur4},\cite{gur6}. In this case there are $N+1$ number complex fields $\cal{E}_{k}$ ($k=1,2, \cdots , N+1$) satisfying the coupled equations.

\begin{equation}
({\cal{E}}_{k} \bar{{\cal{E}}}_{k}-1)\, \nabla^2 \, {\cal{E}}_{i}=2 \,\bar{{\cal{E}}}_{k}\, (\vec{\nabla} {\cal{E}}_{k}\, \cdot \vec{\nabla} {\cal{E}}_{i}),  ~~~i=1,2, \cdots, N+1 \label{denk40}\\
\end{equation}
Here the repeated indices are summed up from 1 to $N+1$. An immediate local reduction of these equations is given as
${\cal{E}}_{i}=\alpha_{i}+\beta{i}\, E_{0}$ or ${\cal{E}}_{i}=\alpha_{i}+\beta_{i}\, \bar{E}_{0}$,
where $\alpha_{i}$ and $\beta_{i}$ ($i=1,2,\cdots,N+1$) are complex numbers satisfying
$ \alpha_{k} \,\bar{\beta}_{k}=0$ and $\alpha_{k} \,\bar{\alpha}_{k}+\beta_{k}\, \bar{\beta}_{k}=1$, for all $N \ge 1$. Here
the complex function $E_{0}$ satisfies the Ernst equation (\ref{ernst}).

\vspace{0.5cm}
\noindent
There are two types of symmetry transformations of the system of equations (\ref{denk40}).

\vspace{0.3cm}
\noindent
(1) Let ${\cal{E}^{\prime}}_{i}=\Lambda^{1}_{ik}\, {\cal{E}}^{k}$ with $\Lambda^{1}\, {\Lambda^{1}}^{\dagger}=I $

\vspace{0.3cm}
\noindent
(2) Let ${\cal{E}^{\prime}}_{i}=\Lambda^{2}_{ik}\, \bar{\cal{E}}^{k}$ with $\Lambda^{2}\, {\Lambda^{2}}^{\dagger}=I$

\vspace{0.5cm}
\noindent
where $\dagger$ denotes Hermitian conjugation and $I$ is the $(N+1) \times (N+1)$ unit matrix . In both types the total number of real parameters is $2 (N+1)^2$.

\vspace{0.5cm}
\noindent
It is possible to write the system equations in a more symmetrical form. In this case we assume that $N$ is an odd integer. Let us define $\xi_{i}={\cal{E}}_{i}$  and $\eta_{i}={\cal{E}}_{i+\frac{N+1}{2}}$. for ($i=1,2,\cdots, \frac{N+1}{2}$). Then the equations (\ref{denk40}) reduce to the following coupled equations
\begin{eqnarray}
&&(\xi_{k} \bar{\xi_{k}}+\eta_{k}\, \bar{\eta_{k}}-1)\, \nabla^2 \, \xi_{i}=2 (\bar{\xi_{k}}\, \vec{\nabla} \xi_{k}+ \bar{\eta}_{k}\, \vec{\nabla} \eta_{k})\, \cdot \vec{\nabla} \xi_{i}, \label{denk4}  \\
&&(\xi_{k} \bar{\xi_{k}}+\eta_{k}\, \bar{\eta_{k}}-1)\, \nabla^2 \, \eta_{i}=2 (\bar{\xi_{k}}\, \vec{\nabla} \xi_{k}+ \bar{\eta_{k}}\, \vec{\nabla} \eta_{k})\, \cdot \vec{\nabla} \eta_{i}, \label{denk5}
\end{eqnarray}
where $i=1,2, \cdots , \frac{N+1}{2}$. $N=1$ corresponds to EM Ernst equations.
This system has similar local and nonlocal reductions.

\vspace{0.5cm}
\noindent
{\bf Local Reductions}:

\vspace{0.5cm}
\noindent
{\bf (1)}. Let $\eta^{\prime}_{k}= a \Lambda_{k \ell}\, \xi_{\ell}$ where $\Lambda$ is a unitary matrix and $a$ is a constant, then the above equations reduce to the following one

\begin{equation}
((1+a \bar{a})\,\xi_{k} \bar{\xi_{k}}-1)\, \nabla^2 \, \xi_{i}=2 ((1+a \bar{a})\,\bar{\xi}_{k}\, \vec{\nabla} \xi_{k} \, \cdot \vec{\nabla} \xi_{i}, \label{denk6}
\end{equation}
for $k=1,2, \cdots , \frac{N+1}{2}$.

\vspace{0.5cm}
\noindent
{\bf (2)}. Let $\eta^{\prime}_{k}= a \Lambda_{k \ell}\, \bar{\xi}_{\ell}$
where $\Lambda$ is a unitary matrix and $a$ is a constant satsifying $a \bar{a}=1$, then (\ref{denk4}) and (\ref{denk5}) reduce to
\begin{equation}
(2\,\xi_{k} \bar{\xi_{k}}-1)\, \nabla^2 \, \xi_{i}=2 \, \vec{\nabla} (\xi_{k} \, \bar{\xi}_{k})\cdot \vec{\nabla} \xi_{i}, \label{denk7}
\end{equation}
for $k=1,2, \cdots , \frac{N+1}{2}$. This equation generalizes the equation (\ref{eq1})

\begin{equation}
\vec{\nabla} \cdot \left( \frac{{\vec{\nabla}} \xi_{i}} {2 \xi_{k} \,\bar{\xi}_{k}-1} \right)=0 , ~~(i=1,2, \cdots , \frac{N+1}{2}) \label{eq2}
\end{equation}

\vspace{0.5cm}
\noindent
{\bf Nonlocal Reductions}:  Let
\begin{equation}
\xi^{\varepsilon}=\xi(\varepsilon_{1} x, \varepsilon_{2} y), ~~\eta^{\varepsilon}=\eta(\varepsilon_{1} x, \varepsilon_{2} y), ~~~(\varepsilon_{1})^2=(\varepsilon_{2})^2=1
\end{equation}

\vspace{0.5cm}
\noindent
{\bf (1)}.  Let $\eta_{i}(x,y)=k \, \xi_{i}(\varepsilon_{1} x, \varepsilon_{2} y)$. Then \ref{denk4}) and (\ref{denk5}) consistently reduce to

\begin{equation}
(\xi_{i} \bar{\xi_{i}}+\xi_{i}^{\varepsilon}\, \bar{\xi}_{i}^{\varepsilon}-1)\, \nabla^2 \, \xi_{k}=2 (\bar{\xi_{i}}\, \vec{\nabla} \xi_{i}+ \bar{\xi_{i}^{\varepsilon}}\, \vec{\nabla} \xi_{i}^{\varepsilon})\, \cdot \vec{\nabla} \xi_{k}, \label{denk10}
\end{equation}
for $k=1,2, \cdots , \frac{N+1}{2}$.

\vspace{0.5cm}
\noindent
{\bf (2)}.  Let ~~$\eta_{k}(x,y)=k \, \bar{\xi_{k}}(\varepsilon_{1} x, \varepsilon_{2} y)$. Then Then \ref{denk4}) and (\ref{denk5}) reduce to

\begin{equation}
(\xi_{i} \bar{\xi_{i}}+\xi_{i}^{\varepsilon}\, \bar{\xi_{i}}^{\varepsilon}-1)\, \nabla^2 \, \xi_{k}=2 (\bar{\xi_{i}}\, \vec{\nabla} \xi_{i}+ \xi_{i}^{\varepsilon}\, \vec{\nabla} \bar{\xi_{i}}^{\varepsilon})\, \cdot \vec{\nabla} \xi_{k}, \label{denk11}
\end{equation}
for $k=1,2, \cdots , \frac{N+1}{2}$, Here $k \bar{k}=1$ for both cases. These are vectorial nonlocal Ernst equations.

\section{Reflection and Shifted Reflection Symmetry}

In the study of the asymptotically flat , stationary axially symmetric solutions of the Einstein field equations it was shown that there exists a class exact solutions of such  configurations possessing reflection symmetry \cite{kordas}- \cite{ernst4}. In such a class the Ernst potential satisfies $\xi(\rho,-z)=\bar{\xi}(\rho,z)$ in cylindrical coordinates. Reflection symmetry for the electro vacuum field equations was studied by Ernst et al \cite{ernst3}, \cite{ernst4}. Here we verify this fact also in the stationary axially symmetric Einstein Maxwell field equations (SASEM). We showed that, in nonlocal reductions, there are two nonlocal Ernst equations representing the (SASEM) which are given in (\ref{eq4}) and (\ref{eqn5}). Although the reflection symmetry was defined in \cite{kordas}-\cite{ernst4} in cylindrical coordinates we shall present our results in prolate spheroidal coordinates. We can easily convert our results to cylindrical coordinates. We have the following Lemma:

\vspace{0.5cm}
\noindent
{\bf Lemma}: {\it SASEM equations (\ref{denk1})-(\ref{denk2}) have a subclass of solutions possessing the reflection symmetry}

\vspace{0.5cm}
\noindent

\begin{equation}
\xi(\varepsilon_{1} x, \varepsilon_{2} y)=k\, \bar{\xi}(x,y)
\end{equation}
where $\varepsilon_{1}^2=\varepsilon_{2}^2=1$ and $k \bar{k}=1$. If we assume the above reflection symmetry the nonlocal Ernst equation (\ref{eq4}) reduces to the local equation (\ref{eq1}) and nonlocal Ernst equation (\ref{eqn5}) reduces to the local Ernst equation. Hence such a reflection symmetric class is consistent and exist. Here asymptotical flatness is not required.

The above Lemma  can  extended also to N-abelian equations. There exist reflection symmetric solutions of the stationary axially symmetric Einstein-N-Abelian Yang-Milles field equations. In this case  the reflection symmetry is given as 

\begin{equation}
\xi_{i}(\varepsilon_{1} x, \varepsilon_{2} y)=k\, \bar{\xi}_{i} (x,y)
\end{equation}
for $i=1,2, \cdots , \frac{N+1}{2}$, in the nonlocal vector Ernst equations (\ref{denk10}) and (\ref{denk11}). Here $ k \bar{k}=1$.
In all these cases asymptotically flatness is is not required.

\vspace{0.5cm}
\noindent
Reflection symmetry is more transparent when  the cylindrical coordinates $\rho$ and $z$ are used, where
\[
\rho=(x^2-1)^{1/2}\,(1-y^2)^{1/2},~~~~z=xy
\]
In this case the differential operators are given as follows

\begin{eqnarray}
&&\nabla^2 \Psi=\frac{1}{\rho} \frac{\partial}{\partial \rho}\left( \rho \frac{\partial \Psi}{\partial \rho} \right)
+\frac{\partial}{\partial z} \left(\frac{\partial \Psi}{\partial z}\right) , \\
&& \vec{\nabla} \Psi \cdot \vec{\nabla} \Phi= \frac{\partial \Psi}{\partial \rho}\, \frac{\partial \Phi}{\partial \rho}+ \frac{\partial \Psi}{\partial z}\, \frac{\partial \Phi}{\partial z}
\end{eqnarray}
for any functions $\Psi$ and $\Phi$.

\vspace{0.5cm}
\noindent
We can show that
$$\eta(\rho,z)=k \bar{\xi}(\rho,-z)$$
 is a consistent nonlocal reduction of the equations (\ref{denk1}) and (\ref{denk2}) where $k \bar{k}=1$.
 Then the nonlocal Ernst equation in cylindrical coordinates takes the form

 \begin{equation}\label{eqn15}
(\xi \bar{\xi}+\xi^{\varepsilon}\, \bar{\xi}^{\varepsilon}-1)\, \nabla^2 \, \xi=2 (\bar{\xi}\, \vec{\nabla} \xi+ \xi^{\varepsilon}\, \vec{\nabla} \bar{\xi}^{\varepsilon})\, \cdot \vec{\nabla} \xi,
\end{equation}
where $\xi^{\varepsilon}=\xi(\rho,-z)$ in this case. Then the reflection symmetric solutions in these coordinates satisfy $\xi(\rho,z)=k \bar{\xi}(\rho,-z)$, \cite{kordas}-\cite{ernst4}. In prolate spheroidal coordinates $x$ and $y$ the reflection symmetry implies $\xi(x,y)=k \bar{\xi}(x,-y)$ for vacuum field equations. The Kerr solution  $\xi(x,y)=p x+i qy$ has this reflection symmetry where $p$ and $q$ are constants related to the mass and spin. Similarly the Kerr-Newmann solution has also this symmetry.

\vspace{0.5cm}
\noindent
The reflection symmetry defined above is with respect to the plane $z=0$. It is possible to define a shifted shifted reflection symmetry. Recently Ablowitz and Musslimani have introduced a new reduction called shifted nonlocal reduction \cite{ab-muss4}. According to this reduction we let $\eta(\rho,z)=k \bar{\xi}(\rho,z_{0}-z)$ where $z_{0}$ is any real constant. Then the system of equations (\ref{denk1}) and (\ref{denk2}) reduce to a single nonlocal (shifted nonlocal) equation (\ref{eqn15}) where $\xi^{\varepsilon}=\xi(\rho,z_{0}-z)$. It reads

 \begin{eqnarray}\label{eqn16}
&&\left(\xi(\rho,z) \bar{\xi}(\rho,z)+\xi(\rho, z_{0}-z)\, \bar{\xi}(\rho,z_{0}-z)-1 \right)\, \nabla^2 \, \xi(\rho,z) \nonumber\\
&&=2 \left(\bar{\xi}(\rho,,z)\, \vec{\nabla} \xi(\rho,z)+ \xi(\rho,z_{0}-z)\, \vec{\nabla} \bar{\xi}(\rho,z_{0}-z)\right)\, \cdot \vec{\nabla} \xi,
\end{eqnarray}

This is the third nonlocal Ernst equation presented in this work,
We then define a shifted reflection symmetric solutions as $\xi(\rho,z)=k \bar{\xi}(\rho,z_{0}-z)$. The Lemma  can be extended for such reflected symmetric solutions. We claim that shifted reflection symmetry of the Einstein-Maxwell field equations may have some important physical applications.

%\newpage
\section{Conclusion}

 We have obtained nonlocal reductions of the Einstein-Maxwell Ernst equations and also the generalization of these equations, Einstein-N-abelian Yang-Mills Equations. We presented our results for the stationary axially symmetric spacetimes but similar reductions exist also for the colliding plane gravitational plane wave spacetimes.

\vspace{0.3cm}
\noindent
By the use of the nonlocal Ernst equations we showed the existence of a class of solutions of SASEM field equations possessing  reflection symmetry. Reflection symmetry which has been studied by \cite{kordas}-\cite{ernst4} is intimately related to the nonlocal reductions to the Ernst equations. Recently Ablowitz and Musslimani have introduce a new consistent reduction in integrable systems \cite{ab-muss4}. This reduction induces a new symmetry, shifted reflection symmetry. This symmetry is new and needs to be explored and studied in vacuum and electrovacuum Einstein field equations.

\end{document}